\documentclass[runningheads]{llncs}

\usepackage{graphicx}
\usepackage{hyperref}

\usepackage{xcolor}

\usepackage{amsfonts}
\usepackage{amsmath}
\usepackage{mathtools}
\usepackage{wasysym}

\newcommand{\dom}[1]{\mathbb{#1}}

\newcommand{\ad}[1]{\mathbf{a}_{#1}}
\newcommand{\ada}[2]{\mathbf{a}^{#1}_{#2}}

\newcommand{\bmd}[1]{\mathbf{b}_{#1}}
\newcommand{\bmda}[2]{\mathbf{b}^{#1}_{#2}}

\newcommand{\ux}[1]{u_{#1}}

\newcommand{\ppa}[2]{\mathbf{P}^{#1}_{#2}}

\newcommand{\opia}[2]{\omega^{#1}_{#2}}

\newcommand{\qm}[1]{``#1''}

\begin{document}

\title{Opinion Update in a Subjective Logic Model for Social Networks}

\author{Mário S. Alvim\inst{1} \and
Sophia Knight\inst{2} \and
José C. Oliveira\inst{2}}

\authorrunning{M. S. Alvim et al.}

\institute{Department of Computer Science, UFMG, Brazil \\ \email{msalvim@dcc.ufmg.br} \and
Department of Computer Science, University of Minnesota Duluth, USA \email{sophia.knight@gmail.com, josecarlosdeoliveirajr@gmail.com}
}

\maketitle

\begin{abstract}
\emph{Subjective Logic} (SL) is a logic incorporating uncertainty and opinions for agents in dynamic systems. In this work, we investigate the use of subjective logic to model opinions and belief change in social networks. In particular, we work toward the development of a subjective logic belief/opinion update function appropriate for modeling belief change as communication occurs in social networks. We found through experiments that an update function with belief fusion from SL does not have ideal properties to represent a rational update. Even without these properties, we found that an update function with cumulative belief fusion can describe behaviors not explored by the social network model defined by Alvim, Knight, and Valencia \cite{alvim2019toward}.
\end{abstract}

\section{Introduction}

Recently, social networks have begun to influence every aspect of our lives, with unprecedented, unanticipated consequences, especially in politics and public opinion.  Research on social networks has studied opinions and their change over time, but to accurately model real people and their opinions and beliefs, we must include representations of \emph{uncertainty} in formal models of social networks.

To achieve this goal, we use \emph{subjective logic} (SL), which includes information about agents' uncertainty, to develop a more nuanced model of social networks and their changes over time. This work builds upon the model developed by Alvim, Knight, and Valencia (AKV) \cite{alvim2019toward}. Their social network model incorporates quantitative opinions and influence on each agent but only addresses binary opinions and uncertainty is not represented.

The contributions of this paper are the following:
\begin{itemize}
    \item We propose a model for social networks using elements of the subjective logic model such as multinomial opinions, trust opinions, and belief fusion operators.
    
    \item We propose a belief/opinion update function using SL's trust discount and belief fusion. We use examples to show that our update function, with cumulative, averaging, and weighted belief fusions, does not satisfy proprieties that are useful to model a basic social network with rational update.
    
    \item We analyze the update function using trust discount and cumulative belief fusion from subjective logic and how it can represent a different scenario not described in the model proposed by Alvim, Knight, and Valencia \cite{alvim2019toward}.
\end{itemize}

\section{Related Work}

There is a great deal of work concerned with logical models of social networks and formalizing belief change in social networks, but we are unaware of other work modeling quantitative uncertainty in social networks. 
More specifically, other work allows uncertainty only in the sense of multiple states or worlds that an agent considers possible, whereas in this work we use subjective logic,  considering uncertainty as a \emph{total lack of information}, on top of the multiple states an agent may consider possible.
Much of the literature treats opinions as binary, non-quantitative phenomena, whereas we investigate opinions that take on a spectrum of values between 0 and 1, and which also include possible uncertainty.  

Before the advent of online social networks, Degroot et al. \cite{degroot1974reaching} proposed a model of learning and consensus in multi-agent systems, in which quantitative beliefs are updated by a constant stochastic matrix at each time step. The Degroot model does not include uncertainty but otherwise is close to our work in spirit. The models in \cite{bala1998learning} are also similar to the models we use, but the focus of that work is on payoffs and optimal decision-making, whereas our focus is purely on the changes in information and uncertainty over time. Incorporating goals and decisions into our models provides an interesting problem for future work, particularly in the context of our focus on uncertainty. In \cite{holliday2009trust}, Holliday develops a logic with \emph{ordered} but non-quantitative trust and certainty about propositions: agent $a$ may trust agent $b$ more than they trust agent $c$, leading them to believe certain propositions more strongly than others. Liu et al. \cite{liu2014logical} use ideas from doxastic and dynamic epistemic logics to qualitatively model influence and belief changes in social networks. Christoff \cite{christoff2016dynamic} has developed several non-quantitative logics for social networks, and Young Pedersen \cite{PSA,pedersen2019polarization} develops a non-quantitative logic concerned specifically with polarization. 
In \cite{XA}, Xiong and \AA gotnes develop a logic to analyze signed social networks where agents can have ``friends'' and ``enemies,''  a different approach to some of the same questions that concern us, such as polarization and influence.

Hunter \cite{hunter2017reasoning} introduces a logic of belief updates over social networks with varying levels of influence and trust.
Using dynamic epistemic logic, Baltag et al. \cite{baltag2019dynamic} created a threshold model where agents' behavior changes when the proportion of supporters changes, but with binary belief and no uncertainty.

This work is a continuation of Alvim et al.'s work \cite{AKQV21,AKQV,alvim2019toward}. Alvim et al. develop a formal model for social networks where agents have quantitative opinions and quantitative influence on each other, with a function for agents' belief update over time. The goal of the current paper is to extend this model by adding the possibility of \emph{uncertainty} to the agents' quantitative opinions.

\section{Background: Subjective Logic}\label{sec:subjective_logic}

This section provides background on the elements of subjective logic that we use in our model. \emph{Subjective Logic} is a logic developed by J{\o}sang \cite{josang2016subjective} that extends probabilistic logic by adding \emph{uncertainty} and \emph{subjectivity}. 
In probabilistic logic, a uniform distribution does not express \emph{\qm{we don't know}} because a uniform distribution says that we know that the distribution over the domain is uniform.
Subjective logic can distinguish between the situation where the distribution over outcomes is unknown and the situation where the distribution is known and, for example, uniform. In subjective logic, it is also possible to have a situation where some information about the distribution is known and there is some uncertainty. 
The \emph{subjectivity} comes from the fact that we can assign an opinion, or information, about a proposition to an agent.

\subsubsection*{Opinion representation}\label{sec:opinion_representation}

The main object of subjective logic is the \emph{opinion}. 
We represent an opinion by $\opia{A}{X}$, where $A$ is an agent, $X$ a random variable, and $\opia{A}{X}$ is $A$'s opinion about $X$. An opinion expresses support for none, one, or many states of a domain.
This section presents the elementary definitions that compose an opinion.
A \emph{domain} is a state space consisting of a finite set of values called states, events, outcomes, hypotheses, or propositions. The values are assumed to be exclusive and exhaustive.

\emph{Belief mass} is a distribution over a domain $\dom{X}$ representing an agent's confidence in each value in the domain. The belief mass assigned to a value $x \in \dom{X}$ expresses support for $x$ being TRUE. Belief mass is sub-additive, i.e. $\sum\limits_{x \in \mathbb{X}} \mathbf{b}_X(x) \leq 1$. The sub-additivity is complemented by \emph{uncertainty mass} $\ux{X}$ and it represents the lack of support or evidence for the variable $X$ having any specific value. 

\begin{definition}\label{def:belief_mass_distribution}
	\emph{(Belief Mass Distribution)} Let $\dom{X}$ be a domain of size $k \geq 2$, and let $X$ be a variable over that domain. A belief mass distribution denoted $\mathbf{b}_X : \dom{X} \rightarrow [0, 1]$ assigns belief mass to possible values of the variable $X$. Belief mass and uncertainty mass sum to one, i.e., $u_X + \sum\limits_{x \in \mathbb{X}} \mathbf{b}_X(x) = 1$.
\end{definition}

Opinions can be semantically different, depending on the situation they apply to. 
An \emph{aleatory} opinion applies to a variable governed by a random or frequentist 
process, and represents the likelihood of values of the variable in any unknown past or future instance of the process. \qm{\emph{The (biased) coin will land \emph{heads} with $p = 0.6$}} is an aleatory opinion.
An \emph{epistemic} opinion applies to a variable that is assumed to be non-frequentist, and that represents the likelihood of the variables in a specific unknown instance. \qm{\emph{Beatriz killed Evandro}} is an epistemic opinion.
In an epistemic opinion, opposite/different pieces of evidence should cancel each other out. Therefore, it must be uncertainty-maximized. 

Base rate distribution represents a \emph{prior} probability distribution over a domain: the probability distribution before considering evidence about the domain.

\begin{definition}\label{def:opinion}
	\emph{(Opinion)} Let $\mathbb{X}$ be a domain of size $k \geq 2$, and $X$ a random variable in $\mathbb{X}$. An opinion over the random variable $X$ is the ordered triple $\omega_X = (\mathbf{b}_X, u_X , \mathbf{a}_X)$ where
	\begin{itemize}
		\item $\mathbf{b}_X$ is a belief mass distribution over $X$,
		\item $u_X$ is the uncertainty mass which represents a lack of evidence,
		\item $\mathbf{a}_X$ is a base rate distribution (a probability distribution) over $\mathbb{X}$.
	\end{itemize}
\end{definition}

The \emph{projected probability distribution} 
of an opinion is the \emph{posterior} probability distribution after updating the base rate distribution with the belief mass distribution.
The more an opinion depends on the belief mass, the less it depends on the base rate. The projected probability distribution is defined by
$    \mathbf{P}_X(x) = \bmd{X}(x) + \ad{X}(x) u_X,\ \forall x \in \dom{X}$

This representation is equivalent to representing an opinion as a Beta PDF (or Dirichlet PDF if $k > 2$), where the posterior probability is obtained by updating the parameters $\alpha$ and $\beta$ (or a vector of parameters $\alpha$ for the Dirichlet PDF) given the observations. The equivalence is defined as the projected probability from SL's opinion being equivalent to the expected probability of the Beta PDF. More details about the equivalence between opinions and Beta PDFs are presented in Appendix \ref{sec:mappin_an_opinion_to_a_beta_pdf}.

The definition of opinion is useful for our model since it is more expressive than the belief state of an agent about a proposition in \cite{alvim2019toward}, which is similarly an opinion with domain $\dom{X} = \{true, false\}$, with no uncertainty mass. The agent must commit all of their mass to the values of the domain with no uncertainty.
\begin{example}
Let $\dom{X} = \{x, \overline{x}\}$ be a domain where $x$ is \qm{global warming is happening} and $\overline{x}$ is \qm{global warming is \emph{not} happening}. 
Let $X$ be a random variable in $\dom{X}$. An opinion about $X$ must be epistemic because it is about a fact in the present instance that is true or false.
Let the base rate be uniform. With no evidence, an agent $A$ will hold the opinion $\opia{A}{X} = ((0, 0), 1, (0.5, 0.5))$ with $\ppa{A}{X}(x) = 0.5$, meaning that $A$ is $50\%$ sure that the global warming is happening, but their opinion is relying only on the base rate, with no evidence supporting either of the values.

After gathering evidence from newspapers, scientific studies, and other people, $A$ assigns a belief mass to $x$.
If agent $A$ holds the opinion $\opia{A}{X} = ((0.6, 0), 0.4, \allowbreak (0.5, 0.5))$, then $\ppa{A}{X}(x) = 0.8$. In this case, $A$ is $80\%$ sure that global warming is happening, and has evidence that corresponds to $60\%$ of their mass. The uncertainty mass means that $A$ is relying $40\%$ on the base rate.
\end{example}

\subsubsection*{Trust discount}

To model the influence that one agent has on another, subjective logic has \emph{trust opinion}, an opinion an agent has about another agent as a source of information. 

\begin{definition}
    \emph{(Trust opinion)} Let $\dom{T}_B = \{t_B, \overline{t}_B\}$ be a \emph{trust domain}, where $t_B$ means \qm{$B$ is a good source of information} and $\overline{t}_B$ means \qm{$B$ is not a good source of information}. Then $\opia{A}{t_B}$, or $\opia{A}{B}$ for short, is the (trust) opinion that $A$ has about the trustworthiness of $B$ as a source of information.
\end{definition}

We use trust opinions to model an agent's updated opinion after communication:  $\opia{[A;B]}{X}$ is a new opinion generated by taking belief mass $\opia{B}{X}$ proportional to the belief mass of the trust opinion $\opia{A}{B}$. And $\opia{[A;B]}{X}$ represents $A$'s opinion about $X$ after communicating with $B$, $\opia{A}{B}$ represents $A$'s opinion about $B$'s trustworthiness, and $\opia{B}{X}$ represents $B$'s opinion about $X$. The operation is denoted  $\opia{[A;B]}{X} = \opia{A}{B} \otimes \opia{B}{X}$. 

There are several options for computing this value, depending on the situation being modeled. Developing an accurate function for opinion updates in social networks is a focus of the current paper.

\begin{example}
Let $\opia{A}{B} = ((1, 0), 0, \mathbf{a}^A_B)$ with $\ppa{A}{B}(t_B) = 1$ and $\opia{B}{X} = ((0.6, 0), 0.4, \linebreak[1] (0.5, 0.5))$ with $\ppa{B}{X}(x) = 0.8$. Here, $A$ completely trusts $B$ and $B$ is $80\%$ sure that $x$ is true with $60\%$ of their mass assigned to $x$.

$\ppa{A}{B}(t_B) = 1$, i.e. $A$ completely trusts $B$. Then, $A$ by trusting $B$ (in short $[A;B]$) will hold the same opinion as $B$ about $X$. Therefore, $\opia{[A;B]}{X} = \opia{B}{X}$. By the opinion that $A$ has about $X$ by trusting $B$, $A$ is $60\%$ sure that $x$ is true with $80\%$ of their mass assigned to $x$.
\end{example}

\begin{example}
Let $\opia{A}{B} = ((0.5, 0.5), 0, \mathbf{a}^A_B)$ with $\ppa{A}{B}(t_B) = 0.5$ and $\opia{B}{X} = ((0.8, 0), \allowbreak 0.2, \linebreak[1] (0.5, 0.5))$ with $\ppa{A}{X}(x) = 0.9$. Here, $A$ trusts $B$ by $50\%$ and $B$ is $80\%$ sure that $x$ is true with $60\%$ of their mass assigned to $x$.

$\ppa{A}{B}(t_B) = 0.5$. Then, $[A;B]$ will hold $50\%$ of the belief mass of each value from $B$. Therefore, $\opia{[A;B]}{X} = ((0.4, 0), 0.6, (0.5, 0.5))$ with $\ppa{[A;B]}{X}(x) = 0.7$. By the opinion that $A$ has about $X$ by trusting $B$, $A$ is $70\%$ sure that $x$ is true with $40\%$ of their mass assigned to $x$.
\end{example}

\subsubsection*{Belief fusion}

To model $A$'s concurrent interactions with multiple other agents, we use belief fusion \cite{josang2016subjective,josang2018categories}. Belief fusion combines a set of opinions into a single opinion which then represents the opinion of the collection of sources. 
There is more than one possible definition for the belief fusion operator. The possible definitions differ in their properties and applications. Below, we omit the details of the calculation of the fusion operators and instead, explain the intuition. For more details, see \cite{josang2016subjective}. For our model, we consider the following operators from \cite{josang2016subjective,josang2018categories}:
\begin{itemize}
    \item \emph{Cumulative belief fusion} (denoted $\opia{(A \diamond B)}{X} = \opia{A}{X} \oplus \opia{B}{X}$): Used when it is assumed that the amount of independent evidence increases by including more sources. The idea is to sum the amount of evidence of the opinions. It is non-idempotent. E.g.\! a set of agents flips a coin several times and produces an opinion about the bias of the coin. An opinion produced by cumulative belief fusion represents all the experiments made by the agents.
    
    \item \emph{Averaging belief fusion} (denoted $\opia{(A \underline \diamond B)}{X} = \opia{A}{X} \underline \oplus \opia{B}{X}$): It is used when including more sources does not mean that more evidence supports the conclusion. The idea is to take the average of the amount of evidence of the opinions. It is idempotent, but it has no neutral element.
    E.g. After observing the court proceedings, each member of a jury produces an opinion about the same evidence. The verdict is the fusion between those opinions.
    
    \item \emph{Weighted belief fusion} (denoted $\opia{(A \hat \diamond B)}{X} = \opia{A}{X} \hat \oplus \opia{B}{X}$): It is used when we take the average of the amount of evidence of the opinions weighted by their lack of uncertainty. In particular, opinions with no belief mass are rejected. It is idempotent and it has a neutral element $u_X = 1$.
    E.g. a group of medical doctors needs to decide on a diagnosis. Each of them has an opinion, but some of them are more certain (assigned more belief mass) than others. Those opinions must have more weight than the others upon fusion.
\end{itemize}

Each of these operators is defined in the Appendix \ref{sec:belief_fusion_operators}.

\section{A Subjective Logic model for social networks}

We describe how to use subjective logic to model a social network. Existing notions from subjective logic work well for modeling the relevant aspects of agents' opinions about an issue in a social network, except for the update function. Our goal is to expand subjective logic with an appropriate update function. 

\subsection{Static elements of the model}

The \emph{static elements} represent a snapshot of a social network at a given point in time. They include the following components:
\begin{itemize}
    \item A (finite) set $\mathcal{A} = \{A_0, A_1, \dots, A_{n-1}\}$ of $n \geq 1$ \emph{agents}. An agent is a user in a social network

    \item A \emph{domain} of $k$ disjoint events $\dom{X} = \{x_0, x_1, \dots, x_{k-1}\}$ and a random variable $X$ over $\dom{X}$. A domain is a generalization of a proposition from binary logic by having multiple truth values for a proposition. A proposition from binary logic would have a domain of size $2$. It can represent topics that cannot be answered with just a YES or NO. In the examples in this work, we use a domain of size two for simplicity, and the random variable $X$ represents opinions about a single issue.

    \item A set of \emph{opinions} $\{\opia{A_j}{X}\}_{A_i \in \mathcal{A}}$, one for each agent, about a single random variable $X$.
\end{itemize}

\subsection{Dynamic elements of the model}

The \emph{dynamic elements} of the model formalize the evolution of agents' beliefs as they interact and communicate about their opinions over time. They include:
\begin{itemize}
    \item A set of \emph{trust opinions} $\{\opia{A_i}{A_j}\}_{A_i, A_j \in \mathcal{A}}$ over the domain $\dom{T}_{A_j} = \{t_{A_j}, \overline{t}_{a_j}\}$ where $A_i, A_j \in \mathcal{A}$ and $i\neq j$. Each trust opinion represents how much $A_i$ trusts $A_j$ as a source of evidence. We consider trust opinion as dogmatic, i.e. the uncertainty mass is $i_{T_j} = 0$ for all trust opinions. Therefore, $\ppa{A_i}{A_j}(t_{A_j}) = \bmda{A_i}{A_j}(t_{A_j})$. Throughout this paper, we represent a trust opinion only by $\ppa{A_i}{A_j}(t_{A_j})$.

    \item A \emph{time frame} $\mathcal{T} = \{0, 1, 2, \dots, t_{max}\}$ representing the discrete passage of time.

    \item An \emph{update function} $f$ such that
    \begin{equation}
        \opia{A_i[t+1]}{X} = f(\opia{A_i[t]}{X}, \{\opia{A_i}{A_j}\}_{A_j \in \mathcal{A}}, \{\opia{A_j[t]}{X}\}_{A_j \in \mathcal{A}}).
    \end{equation}

    An update function $f$ takes $A_i$'s opinions at time $t$ and updates it using all other agents' opinions and trust opinions to them. Here we consider that agents have complete trust in themselves. Otherwise, we can define this function with $A_i$'s opinion included in the set of opinions $\{\opia{A_i}{A_j}\}_{A_j \in \mathcal{A}}$.
\end{itemize}

\section{Applying trust discount and belief fusion}

The intuition behind our planned update function is to fuse agent $A$'s current opinion with all the opinions that $A$ can gather by trusting other agents. Define a \emph{dogmatic opinion} as an opinion with no uncertainty, i.e. $u_X = 0$. For this update function, we are not considering situations with dogmatic opinions, because it means the agent has an infinite amount of evidence and the belief fusion operators remove non-dogmatic opinions when at least one is present.

\begin{definition}
    \emph{(Update function with Belief Fusion and Trust)}
    Let $\opia{A_0[t]}{X}, \cdots, \allowbreak \opia{A_{n - 1}[t]}{X}$ be non-dogmatic opinions. Let $\oplus$ be a belief fusion operator. Define the update function for $\opia{A_n[t + 1]}{X}$ as
    \begin{equation}
        \opia{A_n[t + 1]}{X} = \opia{A[t]}{X} \oplus \bigoplus_{\substack{A_m \in \mathcal{A} \\ m \neq n}} (\opia{A_n}{A_m} \otimes \opia{A_n[t]}{X}).
    \end{equation}
\end{definition}

Our goal in this section is to understand the effects and properties of the proposed update function applied to only two agents. Doing so will properly set a foundation to understand the same effects and properties on a large-scale model. Now we define the update function for two agents with examples and experiments.

\begin{definition}\label{def:belief-trust-update}
\emph{(Update function for two agents with Belief Fusion and Trust)} Let $\opia{A[t]}{X}$ and $\opia{B[t]}{X}$ be non-dogmatic opinions. Let $\oplus$ be a belief fusion operator. Define the update function for $\opia{A[t + 1]}{X}$ as
\begin{equation}
    \opia{A[t + 1]}{X} = \opia{A[t]}{X} \oplus (\opia{A}{B} \otimes \opia{B[t]}{X}).
\end{equation}
\end{definition}

We call $(\opia{A}{B} \otimes \opia{B[t]}{X})$ the opinion that A {will} learn by interacting with $B$. $\opia{A[t + 1]}{X}$ is the opinion that $A$ holds after merging their previous opinion ($\opia{A[t]}{X}$) with the opinion that $A$ learned $(\opia{A}{B} \otimes \opia{B[t]}{X})$.

\begin{example}
For brevity, we write the value of $\ppa{A}{B}(t_B)$ where it should be $\opia{A}{B}$.
If
\begin{equation}
    \begin{array}{llll}
        \opia{A[t]}{X} & = ((0, 0), 1, (0.5, 0.5)), & \quad \ppa{A[t]}{X}(x) & = 0.5 \\
        \opia{A}{B} & = ((0.5, 0.5), 0, \ada{A}{B}), & \quad \ppa{A}{B}(t_B) & = 0.5 \\
        \opia{B[t]}{X} & = ((0.8, 0), 0.2, (0.5, 0.5)))  & \quad \ppa{B[t]}{X}(x) & = 0.9,
    \end{array}
\end{equation}
then
\begin{equation}
    \begin{array}{ll}
        \opia{A[t + 1]}{X} & = ((0, 0), 1, (0.5, 0.5)) \oplus (0.5 \otimes ((0.8, 0), 0.2, (0.5, 0.5))) \\
        & = ((0, 0), 1, (0.5, 0.5)) \oplus ((0.4, 0), 0.6, (0.5, 0.5)).
    \end{array}
\end{equation}

Agent A is $50\%$ sure about $x$ and $B$ is $90\%$ about $x$. Both trust each other by $50\%$. As we show below, there are different outcomes depending on the choice of fusion operator. 
\end{example}

\begin{example} \label{example5}
Now consider these two agents iteratively updating their opinion over time.
\begin{equation}
\begin{array}{rlll}
    \opia{A[0]}{X} & = ((0.2, 0), 0.8, (0.5, 0.5)) & \quad \ppa{A[0]}{X}(x) = 0.6 & \quad \ppa{A}{B}(t_B) = 0.5\\
    \opia{B[0]}{X} & = ((0.8, 0), 0.2, (0.5, 0.5)) & \quad  \ppa{B[0]}{X}(x) = 0.9 & \quad  \ppa{B}{A}(t_A) = 0.5
\end{array}
\end{equation}

Here, initially agent $A$ is $60\%$ sure about $x$ and $B$ is $90\%$ about $x$. Both trust each other by $50\%$. The iterated evolution of $\ppa{A}{X}(x)$ and $\ppa{B}{X}(x)$ over 20 time steps is shown in Fig. \ref{fig1}.

\begin{figure}
 \centering
\includegraphics[width=\textwidth]{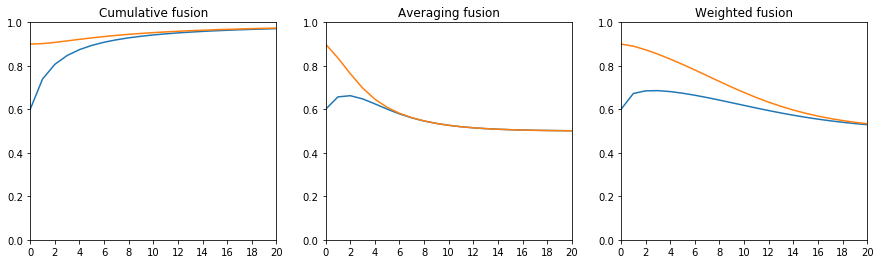}
\caption{$\ppa{A}{X}(x)$ (blue) and $\ppa{B}{X}(x)$ (orange) updated 20 times as in Example \ref{example5}.} \label{fig1}

\end{figure}

Say that weak convergence means that $A$ cannot move further from $B$ upon update, i.e. $\ppa{A[t]}{X} \leq \ppa{A[t + 1]}{X} \leq \ppa{B[t]}{X}$ or $\ppa{A[t]}{X} \geq \ppa{A[t + 1]}{X} \geq \ppa{B[t]}{X}$.
We expected that the update function would weakly converge by $\ppa{A}{X}(x)$ and $\ppa{B}{X}(x)$ converging to some value between $\ppa{A[0]}{X}(x) = 0.6$ and $\ppa{B[0]}{X}(x) = 0.9$.
But because evidence keeps accumulating over time, with cumulative belief fusion, $\ppa{A}{X}(x)$ and $\ppa{B}{X}(x)$ converge to $1$.
Therefore, the update function with cumulative fusion does not weakly converge.

For averaging and weighted belief fusion, $\ppa{A}{X}(x)$ and $\ppa{B}{X}(x)$ converge to $0.5$, violating weak convergence. With epistemic opinions, increasing uncertainty over time is expected, but the same happens with aleatory opinions because the trust discount removes an amount of evidence from each agent at every step.
\end{example}

\begin{example} \label{example7}
In this case, $A$ and $B$ have the same opinion and the same trust.
\begin{equation} 
\begin{array}{rlll}
    \opia{A[0]}{X} & = ((0.6, 0), 0.4, (0.5, 0.5)) & \quad \ppa{A[0]}{X}(x) = 0.8& \quad  \ppa{A}{B}(t_B) = 0.5 \\
    \opia{B[0]}{X} & = ((0.6, 0), 0.4, (0.5, 0.5)) & \quad  \ppa{B[0]}{X}(x) = 0.8 & \quad \ppa{B}{A}(t_A) = 0.5
\end{array}
\end{equation}

Here, $A$ and $B$ have the same opinion. They are $80\%$ sure about $x$. Both trust each other by $50\%$. The evolution of $\ppa{A}{X}(x)$ and $\ppa{B}{X}(x)$ is shown in  Fig. \ref{fig3}.

\begin{figure}
 \centering

\includegraphics[width=\textwidth]{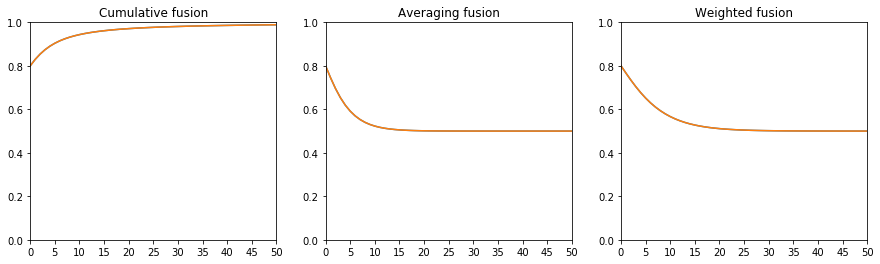}
\caption{$\ppa{A}{X}(x)$ and $\ppa{B}{X}(x)$ (both orange) updated 20 times as in Example \ref{example7}.} \label{fig3}
\end{figure}

Even starting with the same opinion, agents $A$ and $B$ do not keep the same opinion over time. With cumulative belief fusion, $A$ and $B$ keep accumulating evidence and $\ppa{A}{X}(x)$ and $\ppa{B}{X}(x)$ converge to $1$. With averaging or weighted belief fusion, uncertainty keeps increasing, and $\ppa{A}{X}(x)$ and $\ppa{B}{X}(x)$ converge to $0.5$. The example shows none of the belief fusion operators make the update idempotent with respect to the agents' opinions. None of these operators can represent the case where agents with the same opinion, when interacting, keep the same opinion over time.
\end{example}

In this section, we showed that an update function when belief fusion and trust discount do not have ideal properties that can be useful to model a rational update in a social network. Even without these properties, in the next section, we show that an update function with cumulative belief fusion can represent a different kind of bias not explored in Alvim, Knight, and Valencia's work \cite{AKQV21,AKQV,alvim2019toward}.

\section{Analysis of the update function with cumulative belief}

In this section, we analyze the cumulative belief fusion operator, which we believe in practice provides some interesting outcomes in simple simulations of social networks, even though it does not always weakly converge, as seen above.
We will show that the cumulative belief function is promising for modeling belief updates in social networks.

Recalling the definition of cumulative belief fusion, it is assumed that the amount of evidence increases by including more sources. An opinion in subjective logic can be translated to a Beta PDF using the amount of evidence for each state in the domain as parameters. Say that $\mathbf{r}^A_X(x)$ are $\mathbf{r}^A_X(x)$ the amount of evidence supporting the state $x$ for agents $A$ and $B$ respectively. The cumulative fusion operator gets the sum $\mathbf{r}^A_X(x) + \mathbf{r}^A_X(x)$ to be the amount of evidence of the merged opinion and translates it back to a subjective logic opinion. More details are presented in the appendix. For repeated interactions using the cumulative belief fusion, we interpret repeated evidence contained in an opinion as reinforcement of the belief in the state.

We did many simulations using the two-agent update function with cumulative fusion and we found three cases for initial epistemic opinions with the projected probability $\ppa{A}{X}(x)$ ranging from $0$ to $1$.

Suppose that $\dom{X} = \{x, \overline{x}\}$ and all opinions are epistemic. Recall that the update function for two agents with trust discount and belief fusion is
\begin{equation}
    \opia{A[t + 1]}{X} = \opia{A[t]}{X} \oplus (\opia{A}{B} \otimes \opia{B[t]}{X})
\end{equation}
where $A$ and $B$ are agents, $\opia{A[t]}{X}$ and $\opia{B[t]}{X}$ are non-dogmatic epistemic opinions, and $\opia{A}{B}$ is a trust opinion. Also, recall that $\opia{A}{B} \otimes \opia{B[t]}{X}$ is the opinion that $A$ learns after trusting $B$ about $X$. The update function behaves like these three cases:
\begin{enumerate}
    \item \emph{Consensus}: This case happens when both agents are agreeing or disagreeing at the same time, not necessarily with the same projected probability, i.e.
    \begin{equation}
        \begin{array}{c}
            \ppa{A[0]}{X}(x) < 0.5 \text{ and } \ppa{A}{B}(t_b) \otimes \ppa{B[0]}{X}(x) < 0.5 \text{ or} \\
            \ppa{A[0]}{X}(x) > 0.5 \text{ and } \ppa{A}{B}(t_b) \otimes \ppa{B[0]}{X}(x) > 0.5
        \end{array}
    \end{equation}

    When these agents interact, they will accumulate evidence about the same outcome in each interaction. The fusion leads $\bmda{A}{X}(x)$ and $\bmda{B}{X}(x)$ to converge to $0$ or $1$, depending on what they are agreeing upon, and both with uncertainty mass to $0$. Increasing the trust discount or the uncertainty will increase the speed of convergence.

    This represents a situation when two agents agree that $x$ is TRUE, with different levels of projected probabilities. That leads them to be completely certain about the proposition.

    \begin{example} \label{ex:consensus}
        Let $A$ and $B$ be initially in consensus agreeing that $x$ is TRUE as follows.
        \begin{equation}
            \begin{array}{rlll}
                \opia{A[0]}{X} & = ((0.2, 0.0), 0.8, (0.5, 0.5)) & \quad \ppa{A[0]}{X}{x} = 0.6 & \quad \ppa{A}{B}(t_b) = 0.5 \\
                \opia{B[0]}{X} & = ((0.4, 0.0), 0.6, (0.5, 0.5)) & \quad \ppa{B[0]}{X}{x} = 0.7 & \quad \ppa{B}{A}(t_b) = 0.5
            \end{array}
        \end{equation}

        A is $60\%$ sure about $x$ and $B$ is $70\%$ sure about $x$. Both trust each other by $50\%$. On the limit of the interactions between $A$ and $B$, both will be sure by $100\%$ that $x$ is true.
        \begin{equation}
            \ppa{A[t]}{X}(x) = \ppa{B[t]}{X}(x) \xrightarrow[]{t \rightarrow \infty} 1
        \end{equation}

        Similarly if both disagree that $x$ is TRUE, i.e. that $\overline{x}$ is TRUE, $\ppa{A[t]}{X}(x)$ and $\ppa{B[t]}{X}(x)$ will converge to $0$.
        \begin{figure}
            \centering
            \includegraphics[width=0.75\textwidth]{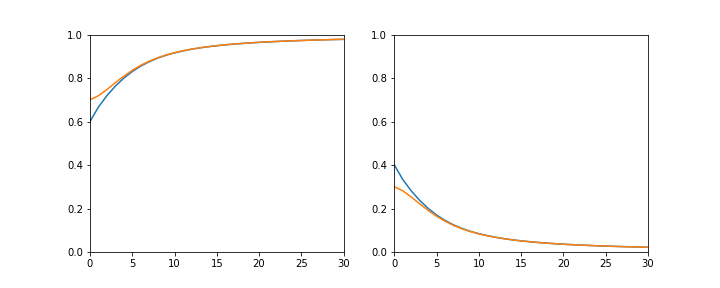}
            \caption{On the left, the evolution of $\ppa{A[t]}{X}(x)$ and $\ppa{A[t]}{X}(x)$ for the Example \ref{ex:consensus}. On the right, the similar case for when both agents disagree that $x$ is TRUE.}
            \label{fig:consensus}
        \end{figure}
    \end{example}

    \item \emph{Balanced opposite}: This case happens when both agents are learning the exact opposite opinion that $A$ already had, i.e.
    \begin{equation}
        \ppa{A[0]}{X}(x) = 1 - \ppa{A}{B}(t_b) \otimes \ppa{B[0]}{X}(x),\ \forall {x} \in \dom{X}
    \end{equation}
    
    Because contrary pieces of evidence cancel each other, the fusion leads $\bmda{A[0]}{X}(x)$ to be a vacuous opinion. The speed of convergence is defined by the trust opinion. The more an agent trusts another, the faster the convergence.
    
    This represents a situation when two agents support opposite views but with the same projected probability. That leads them to be completely indecisive about the proposition.
    
    \begin{example} \label{ex:balance-opposite}
        Let $A$ and $B$ have balanced opposite opinions. 
        \begin{equation}
        \begin{array}{rlll}
            \opia{A[0]}{X} & = ((0, 0.4), 0.6, (0.5, 0.5)) & \quad \ppa{A[0]}{X}(x) = 0.3 & \quad \ppa{A}{B}(t_B) = 1\\
            \opia{B[0]}{X} & = ((0.4, 0), 0.6, (0.5, 0.5)) & \quad  \ppa{B[0]}{X}(x) = 0.7 & \quad  \ppa{B}{A}(t_A) = 1
        \end{array}
        \end{equation}
        
        $A$ is $30\%$ sure about $x$ and $B$ is $70\%$ sure about $x$. Both trust each other by $100\%$. On the limit of interactions between $A$ and $B$, both will be sure by $50\%$ that $x$ is true.
        \begin{equation}
            \ppa{A[t]}{X}(x) = \ppa{B[t]}{X}(x) \xrightarrow[]{t \rightarrow \infty} 0.5
        \end{equation}

        \begin{figure}
            \centering
            \includegraphics[width=0.5\textwidth]{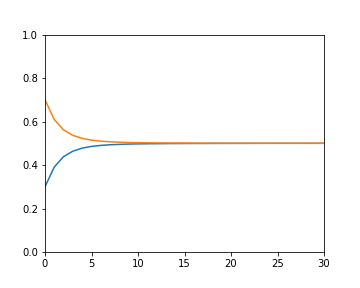}
            \caption{Evolution of $\ppa{A[t]}{X}(x)$ and $\ppa{A[t]}{X}(x)$ for the Example \ref{ex:balance-opposite}.}
            \label{fig:balanced-opposite}
        \end{figure}
    \end{example}

    \item \emph{Unbalanced opposite}: This happens when there are opposite beliefs, but they don't have the same projected probability in their respective views. In other words, both are in conflict but one agent is more radical than the other.
    \begin{equation}
        \begin{array}{c}
            (\ppa{A[0]}{X}(x) < 0.5\ \text{and}\ \ppa{A}{B}(t_b) \oplus \ppa{B}{X} > 0.5 \\ \text{or}\ \ppa{A[0]}{X}(x) > 0.5\ \text{and}\ \ppa{A}{B}(t_b) \oplus \ppa{B}{X} < 0.5) \\
            \text{and}\ \ppa{A[0]}{X} \neq 1 - \ppa{A}{B}(t_b)\ \ppa{B[0]}{X}(x)
        \end{array}
    \end{equation}

    In this case, contrary pieces of evidence will cancel each other, but one agent will transmit more evidence than the other upon interaction. The agents that transmit more evidence will win in the limit. Increasing the trust or becoming more radical will increase the speed of convergence.
    \begin{itemize}
        \item If $A$ is sure that $x$ is TRUE more than $B$ is sure that $x$ is FALSE discounted by the trust, i.e.
        \begin{equation}
            \ppa{A[t]}{X}(x) > 1 - \ppa{A}{B}(t_b) \otimes \ppa{B[t]}{X}(x) 
        \end{equation}
        then, $A$ and $B$ will eventually both agree that $x$ is TRUE in some degree, i.e.
        \begin{equation}
            \ppa{A[t]}{X}(x) = \ppa{B[t]}{X}(x) \xrightarrow[]{t \rightarrow \infty} p \in (0.5, 1]
        \end{equation}

        \item If $A$ is sure that $x$ is FALSE more than $B$ is sure that $x$ is TRUE discounted by the trust, i.e.
        \begin{equation}
            \ppa{A[t]}{X}(x) < 1 - \ppa{A}{B}(t_b) \otimes \ppa{B[t]}{X}(x)
        \end{equation}
        then, $A$ and $B$ will eventually both agree that $x$ is FALSE in some degree, i.e.
        \begin{equation}
            \ppa{A[t]}{X}(x) = \ppa{B[t]}{X}(x) \xrightarrow[]{t \rightarrow \infty} p \in [0, 0.5)
        \end{equation}
    \end{itemize}
\end{enumerate}

The unbalanced opposite is the most interesting case. It can represent some bias when conflicting agents are interacting. The behavior is different from any update function described by \cite{alvim2019toward}. After experiments, we found that
\begin{itemize}
    \item If the agents are too far, they will radicalize, i.e. $\ppa{A[t]}{X}(x) \xrightarrow[]{t \rightarrow \infty} 0$ or \\$\ppa{A[t]}{X}(x) \xrightarrow[]{t \rightarrow \infty} 1$

    \item Otherwise, they will converge to some value at the winning outcome. $[0, 0.5)$ if $A$ is  sure that $x$ is FALSE more than $B$ is  sure that $x$ is TRUE discounted by the trust, or $(0.5, 1]$ if $A$ is  sure that $x$ is TRUE more than $B$ is  sure that $x$ is FALSE discounted by the trust.
\end{itemize}

\begin{figure}
    \centering
    \includegraphics[width=0.5\textwidth]{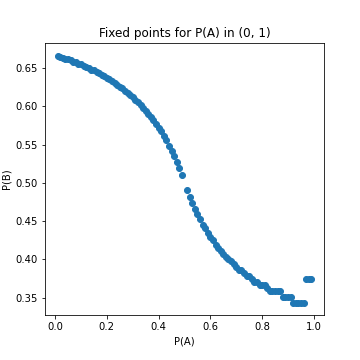}
    \caption{Fixed point for $\ppa{A[0]}{X}(x)$ for each $\ppa{B[t]}{X}(x)$}
    \label{fig:fixed-point}
\end{figure}

This behavior shows that we have a fixed-point for $\ppa{A[0]}{X}(x)$ for each $\ppa{B[t]}{X}(x)$ such that $\ppa{B[0]}{X}(x) = \ppa{B[t]}{X}(x)$ when $t$ goes to infinity, in other words, for every initial $A$'s opinion there is an initial $B$'s opinion such that $A$'s and $B$'s opinions will both eventually converge to $B$'s initial opinion, when $A$ and $B$ communicate repeatedly. 
Fig. \ref{fig:fixed-point} shows the results from the experiments. We found the fixed-points by a binary search for each $\ppa{A[0]}{X}(x)$.

The experiments show that the fixed-point function has a curve similar to the logistic function. 
Each fixed-point represents the boundary between the region where two agents will radicalize and the region where they will eventually agree on some less radical point.
We believe that the radicalization phenomenon happens because the agents are transmitting evidence in an \emph{unstable} way, when the proportion of evidence $x$ vs. $\overline{x}$ increases over time. They do not radicalize when the agents are transmitting evidence in a \emph{stable} way, when the proportion of evidence $x$ vs. $\overline{x}$ is the same over time.

\section{Future Work and Conclusions}
In this work, we described a subjective logic model for social networks. It is defined by a set of agents representing users in a social network, a domain of disjoint events that can be used to represent a topic that cannot be answered with simply YES or NO, a set of multinomial opinions to represent the opinion of each user, trust opinions to represent the relationship between users and an update function to represent the interaction between users.

The main focus of this work is to define an update function. Many update functions can be used depending on what kind of interaction is represented. We did experiments with three update functions using cumulative belief fusion, averaging belief fusion, and weighted belief fusion, with trust discount. 
The experiments showed that none of these functions have useful properties to model the rational update function described in \cite{alvim2019toward} such as idempotency, weak convergence, and non-increasing uncertainty.

Through experiments, we showed that our update function with cumulative belief fusion has the potential to represent a kind of interaction not described in \cite{alvim2019toward}, even though it does not meet our initial desired properties for an update function. We also showed that if the agents disagree about a proposition, i.e. they have opposite opinions, there is a close enough distance between them where the update is stable, and they converge to a non-radical point. But if they are far enough, they will still converge, but also radicalize, converging to completely agree or completely disagree about the proposition. This is different from the update function describing confirmation bias from \cite{alvim2019toward}, where . The next step for the update function with cumulative fusion is to have a clear definition of this behavior by calculating the convergence of the update function for each case.

\bibliographystyle{splncs04}
\bibliography{bibliography}

\appendix

\section{Appendix: Mapping an opinion to a Beta PDF}\label{sec:mappin_an_opinion_to_a_beta_pdf}

This appendix shows the mapping of an opinion to a Beta PDF defined by J{\o}sang \cite{josang2016subjective} We show the evidence notation of the multivariate Beta PDF (or Dirichlet PDF) and show its equivalence with the opinion representation defined at Def. \ref{def:opinion}.

Multinomial probability density over a domain of cardinality $k$ is expressed by the $k$-dimensional Beta PDF. Assume a domain $\dom{X}$ of cardinality $k$, and a random variable $X$ over $\dom{X}$ with probability distribution $\mathbf{p}_X$. The Beta PDF can be used to represent \emph{probability density} over $\mathbf{p}_X$.

The multivariate Beta PDF takes as variable the $k$-dimensional probability distribution $\mathbf{p}_X$. The strength parameters for the $k$ possible outcomes are represented as $k$ positive real numbers $\alpha_X{x}$, each corresponding to one of the possible values $x \in \dom{X}$. The strength parameters represent evidence/observations of $X = x$ where $x \in \dom{X}$.

\begin{definition}
    \emph{(Multivariate Beta Probability Density Function)}. Let $\dom{X}$ be a domain consisting of $k$ mutually disjoint values. Let $\alpha_X$ represent the strength vector over the values of $\dom{X}$, and let $\mathbf{p}_X$ denote the probability distribution over $\dom{X}$. With $\mathbf{p}_X$ as a $k$-dimensional variable, the multivariate Beta PDF denoted $\mathrm{Beta}(\mathbf{p}_X, \alpha_X)$ is expressed as:
    \begin{equation}
        \mathrm{Beta}(\mathbf{p}_X, \alpha_X) = \dfrac{\Gamma\left(\sum\limits_{x \in \mathbb{X}} \alpha_X(x)\right)}{\prod\limits_{x \in \mathbb{X}} \Gamma(\alpha_X(x))} \prod\limits_{x \in \mathbb{X}} \mathbf{p}_X(x)^{(\alpha_X(x)-1)} \text{, where } \alpha_X(x) \geq 0\text{,}
    \end{equation}
    with the restrictions that $\mathbf{p}_X(x) \neq 0$ if $\alpha_X(x)	< 1$.
\end{definition}

Now assume that $x \in \dom{X}$ represents a frequentist event. Let $\mathbf{r}_X(x)$ denote the number of observations for $x$. The strength vector $\alpha_X$ can be expressed as a function of the observations $\mathbf{r}_X(x)$ and the base rate $\mathbf{a}_X$:
\begin{equation}
     \alpha_X(x) = \mathbf{r}_X(x) + \mathbf{a}_X(x)W\text{, where }\mathbf{r}_X(x) \geq 0\ \forall x \in \mathbb{X}\text{.}
\end{equation}

By expressing the strength vector $\alpha_X$ in terms of the evidence observation $\mathbf{r}_X$, the base rate $\mathbf{a}_X$, and the non-informative prior weight $W$, we get the representation of the Beta PDF denoted $\mathrm{Beta}^{\mathrm{e}}_X(\mathbf{p}_X, \allowbreak \mathbf{r}_X, \allowbreak \mathbf{a}_X)$. The exact definition of $\mathrm{Beta}^{\mathrm{e}}_X(\mathbf{p}_X, \allowbreak \mathbf{r}_X, \allowbreak \mathbf{a}_X)$ is described at \cite{josang2016subjective}.

Given a Beta PDF $\mathrm{Beta}^{\mathrm{e}}_X(\mathbf{p}_X, \allowbreak \mathbf{r}_X, \allowbreak \mathbf{a}_X)$, the expected distribution over $\mathbb{X}$ can be written as
\begin{equation}\label{eq:dirithlet_expected_probability}
    \mathbf{E}_X(x) = \dfrac{\mathbf{r}_X(x) + \mathbf{a}_X(x)W}{W + \sum\limits_{x_j \in \mathbb{X}} \mathbf{r}_X(x_j)},\ \forall x \in \mathbb{X}.
\end{equation}

The Beta model translates observation evidence directly into a PDF over a $k$-component probability variable. The representation evidence, together with the base rate, can be used to determine subjective opinions. In other words, it is possible to define a bijective mapping between Beta PDFs and opinions.

The bijective mapping between $\opia{}{X}$ and $\mathrm{Beta}^{\mathrm{e}}_X(\mathbf{p}_X, \allowbreak \mathbf{r}_X, \allowbreak \mathbf{a}_X)$ is based on the requirement for equality between the projected probability distribution $\ppa{}{X}$ derived from $\opia{}{X}$ and the expected probability distribution $\mathbf{E}_X$ derived from \allowbreak $\mathrm{Beta}^{\mathrm{e}}_X(\mathbf{p}_X, \allowbreak \mathbf{r}_X, \allowbreak \mathbf{a}_X)$. This means that the more evidence in favor of a particular outcome $x$, the greater the belief mass on the outcome. Furthermore, the more total evidence available, the less uncertainty mass.

\begin{definition}
    \emph{(Mapping: Opinion $\leftrightarrow$ Beta PDF)} Let $\omega_X = (\mathbf{b}_X, u_X, \mathbf{a}_X)$ be an opinion and let $\mathrm{Beta}^\mathrm{e}_X(\mathbf{p}_X, \mathbf{r}_X, \mathbf{a}_X)$ be a Beta PDF, both over the same variable $X \in \mathbb{X}$. These are equivalent through the following mapping,
    \begin{equation}
        \begin{split}
            & \forall x \in \mathbb{X} \\
            & \begin{cases}
                \mathbf{b}_X(x) & = \dfrac{\mathbf{r}_X(x)}{W + \sum\limits_{x_i \in \mathbb{X}} \mathbf{r}_X(x_i)} \\
                u_X & = \dfrac{W}{W + \sum\limits_{x_i \in \mathbb{X}} \mathbf{r}_X(x_i)}
            \end{cases} \Leftrightarrow
            \begin{cases}
                \begin{cases}
                    \mathbf{r}_X(x) = \dfrac{W \mathbf{b}_X(x)}{u_X} \\
                    1 = u_X + \sum\limits_{x_i \in \mathbb{X}} \mathbf{b}_X(x_i)
                \end{cases} & \text{if } u_X \neq 0 \\
                \begin{cases}
                    \mathbf{r}_X(x) = \mathbf{b}_X(x) \cdot \infty \\
                    1 = \sum\limits_{x_i \in \mathbb{X}} \mathbf{b}_X(x_i)
                \end{cases} & \text{if } u_X = 0
            \end{cases}
        \end{split}
    \end{equation}
\end{definition}

This equivalence between opinions and Beta PDFs is very powerful because it makes it possible to determine opinions from statistical observations.

\section{Appendix: Belief fusion operators}\label{sec:belief_fusion_operators}

This appendix shows a more detailed definition of the belief fusion operators intuitively described at Sec. \ref{sec:subjective_logic}. Here, we will define the belief fusions in terms of Beta PDFs. Since there is a mapping between opinions and Beta PDFs, the direct definition for each operator will be omitted. These definitions can be found at \cite{josang2016subjective,josang2018categories}.

\subsubsection*{Cumulative belief fusion}
The cumulative belief fusion is used when it is assumed that the amount of independent evidence increases by including more sources. The idea is to sum the amount of evidence of the opinions.

Let $\dom{X}$ be a domain, and $X$ be a random variable over $\dom{X}$. W.l.o.g., let $A$ and $B$ be agents, and $\opia{A}{X}$ and $\opia{B}{X}$ be opinions.
The cumulative belief fusion between $A$ and $B$ is denoted $\opia{(A \diamond B)}{X}$ and it can be represented as a Beta PDF.

\begin{definition}
    \emph{(Cumulative Belief Fusion)} Let $\dom{X}$ be a domain, and $X$ be a random variable over $\dom{X}$. Let $A$ and $B$ be agents, and $\opia{A}{X}$ and $\opia{B}{X}$ be opinions. The cumulative belief fusion between $\opia{A}{X}$ and $\opia{B}{X}$ is denoted:
    \begin{equation}
        \opia{(A \diamond B)}{X} = \opia{A}{X} \oplus \opia{B}{X}
    \end{equation}
    
    Let $\mathrm{Beta}^\mathrm{e}_X(\mathbf{p}_X, \mathbf{r}^A_X, \mathbf{a}^A_X)$ and $\mathrm{Beta}^\mathrm{e}_X(\mathbf{p}_X, \mathbf{r}^B_X, \mathbf{a}^B_X)$ be the Beta PDFs equivalent to $\opia{A}{X}$ and $\opia{B}{X}$, respectively.
    
    The opinion $\opia{(A \diamond B)}{X}$ is the opinion equivalent to $\mathrm{Beta}^\mathrm{e}_X(\mathbf{p}_X, \allowbreak \mathbf{r}^{(A \diamond B)}_X, \allowbreak \mathbf{a}^{(A \diamond B)}_X)$ defined as:
    \begin{equation}
        \begin{array}{rl}
            \mathrm{Beta}^\mathrm{e}_X(\mathbf{p}_X, \mathbf{r}^{(A \diamond B)}_X, \mathbf{a}^{(A \diamond B)}_X) & = \mathrm{Beta}^\mathrm{e}_X(\mathbf{p}_X, \mathbf{r}^A_X, \mathbf{a}^A_X) \oplus \mathrm{Beta}^\mathrm{e}_X(\mathbf{p}_X, \mathbf{r}^B_X, \mathbf{a}^B_X) \\
            & = \mathrm{Beta}^\mathrm{e}_X(\mathbf{p}_X, (\mathbf{r}^A_X + \mathbf{r}^B_X), \mathbf{a}^{(A \diamond B)}_X).
        \end{array}
    \end{equation}

    More specifically, for each values $x \in \dom{X}$ the accumulated source evidence $\mathbf{r}^{(A \diamond B)}$ is computed as:
    \begin{equation}
        \mathbf{r}^{(A \diamond B)}(x) = \mathbf{r}^A_X(x) + \mathbf{r}^B_X(x),\ \forall x \in \mathbb{X}.
    \end{equation}
\end{definition}

The fusion of three or more opinions is defined at \cite{josang2018categories}. It can be verified that the cumulative fusion operator is commutative, associative, and non-idempotent.

\subsubsection*{Averaging Belief Fusion}
The averaging belief fusion is used when including more sources does not mean that more evidence supports the conclusion. The idea is to take the average of the amount of evidence of the opinions.

Let $\dom{X}$ be a domain, and $X$ be a random variable over $\dom{X}$. W.l.o.g., let $A$ and $B$ be agents, and $\opia{A}{X}$ and $\opia{B}{X}$ be opinions. The averaging belief fusion between $A$ and $B$ is denoted $\opia{(A \underline \diamond B)}{X} = \opia{A}{X}$ and it can be represented as a Beta PDF.

\begin{definition}
    \emph{(Averaging Belief Fusion)} Let $\dom{X}$ be a domain, and $X$ be a random variable over $\dom{X}$. Let $A$ and $B$ be agents, and $\opia{A}{X}$ and $\opia{B}{X}$ be opinions. The averaging belief fusion between $\opia{A}{X}$ and $\opia{B}{X}$ is denoted:
    \begin{equation}
        \opia{(A \underline \diamond B)}{X} = \opia{A}{X} \underline \oplus \opia{B}{X}
    \end{equation}
    
    Let $\mathrm{Beta}^\mathrm{e}_X(\mathbf{p}_X, \mathbf{r}^A_X, \mathbf{a}^A_X)$ and $\mathrm{Beta}^\mathrm{e}_X(\mathbf{p}_X, \mathbf{r}^B_X, \mathbf{a}^B_X)$ be the Beta PDFs equivalent to $\opia{A}{X}$ and $\opia{B}{X}$, respectively.
    
    The opinion $\opia{(A \underline \diamond B)}{X}$ is the opinion equivalent to $\mathrm{Beta}^\mathrm{e}_X(\mathbf{p}_X, \allowbreak \mathbf{r}^{(A \underline \diamond B)}_X, \allowbreak \mathbf{a}^{(A \underline \diamond B)}_X)$ defined as:
    \begin{equation}
        \begin{array}{rl}
            \mathrm{Beta}^\mathrm{e}_X(\mathbf{p}_X, \mathbf{r}^{(A \underline \diamond B)}_X, \mathbf{a}^{(A \underline \diamond B)}_X) & = \mathrm{Beta}^\mathrm{e}_X(\mathbf{p}_X, \mathbf{r}^A_X, \mathbf{a}^A_X) \underline \oplus \mathrm{Beta}^\mathrm{e}_X(\mathbf{p}_X, \mathbf{r}^B_X, \mathbf{a}^B_X) \\
            & = \mathrm{Beta}^\mathrm{e}_X(\mathbf{p}_X, (\mathbf{r}^A_X + \mathbf{r}^B_X) / 2, \mathbf{a}^{(A \underline \diamond B)}_X).
        \end{array}
    \end{equation}

    More specifically, for each values $x \in \dom{X}$ the accumulated source evidence $\mathbf{r}^{(A \diamond B)}$ is computed as:
    \begin{equation}
        \mathbf{r}^{(A \underline \diamond B)}(x) = \dfrac{\mathbf{r}^A_X(x) + \mathbf{r}^B_X(x)}{2},\ \forall x \in \mathbb{X}.
    \end{equation}
\end{definition}

The fusion of three or more opinions is defined at \cite{josang2018categories}. It can be verified that the cumulative fusion operator is commutative, idempotent, but non-associative.

\subsubsection*{Weighted Belief Fusion}
The weighted belief fusion is used when including more sources does not mean that more evidence supports the conclusion. The idea is used when we take the average of the amount of evidence of the opinions weighted by their lack of uncertainty. In particular, opinions with no belief mass are rejected.

Let $\dom{X}$ be a domain, and $X$ be a random variable over $\dom{X}$. W.l.o.g., let $A$ and $B$ be agents, and $\opia{A}{X}$ and $\opia{B}{X}$ be opinions. The averaging belief fusion between $A$ and $B$ is denoted $\opia{(A \hat \diamond B)}{X} = \opia{A}{X}$ and it can be represented as a Beta PDF.

\begin{definition}
    \emph{(Weighted Belief Fusion)} Let $\dom{X}$ be a domain, and $X$ be a random variable over $\dom{X}$. Let $A$ and $B$ be agents, and $\opia{A}{X}$ and $\opia{B}{X}$ be opinions. The averaging belief fusion between $\opia{A}{X}$ and $\opia{B}{X}$ is denoted:
    \begin{equation}
        \opia{(A \hat \diamond B)}{X} = \opia{A}{X} \hat \oplus \opia{B}{X}
    \end{equation}
    
    Let $\mathrm{Beta}^\mathrm{e}_X(\mathbf{p}_X, \mathbf{r}^A_X, \mathbf{a}^A_X)$ and $\mathrm{Beta}^\mathrm{e}_X(\mathbf{p}_X, \mathbf{r}^B_X, \mathbf{a}^B_X)$ be the Beta PDFs equivalent to $\opia{A}{X}$ and $\opia{B}{X}$, respectively. Also, let $c^A_X$ be the \emph{confidence} that $A$ has in their opinion. Formally:
    \begin{equation}
        c^A_X = 1 - \sum^{i}_{x \in \dom{X}} \bmda{A}{X}(x)
    \end{equation}
    
    The opinion $\opia{(A \hat \diamond B)}{X}$ is the opinion equivalent to $\mathrm{Beta}^\mathrm{e}_X(\mathbf{p}_X, \allowbreak \mathbf{r}^{(A \hat \diamond B)}_X, \allowbreak \mathbf{a}^{(A \hat \diamond B)}_X)$ defined as:
    \begin{equation}
        \begin{array}{rl}
            \mathrm{Beta}^\mathrm{e}_X(\mathbf{p}_X, \mathbf{r}^{(A \hat \diamond B)}_X, \mathbf{a}^{(A \hat \diamond B)}_X) & = \mathrm{Beta}^\mathrm{e}_X(\mathbf{p}_X, \mathbf{r}^A_X, \mathbf{a}^A_X) \hat \oplus \mathrm{Beta}^\mathrm{e}_X(\mathbf{p}_X, \mathbf{r}^B_X, \mathbf{a}^B_X) \\
            & = \mathrm{Beta}^\mathrm{e}_X(\mathbf{p}_X, (c^A_X \mathbf{r}^A_X + c^B_X \mathbf{r}^B_X) / 2, \mathbf{a}^{(A \hat \diamond B)}_X).
        \end{array}
    \end{equation}

    More specifically, for each values $x \in \dom{X}$ the accumulated source evidence $\mathbf{r}^{(A \diamond B)}$ is computed as:
    \begin{equation}
        \mathbf{r}^{(A \hat \diamond B)}(x) = \dfrac{c^A_X \mathbf{r}^A_X(x) + c^B_X \mathbf{r}^B_X(x)}{c^A_X + c^B_X},\ \forall x \in \mathbb{X}.
    \end{equation}
\end{definition}

The fusion of three or more opinions is defined at \cite{josang2018categories}. It can be verified that the cumulative fusion operator is commutative, idempotent, but non-associative.

\end{document}